\documentclass[showpacs,aps,twocolumn]{revtex4}
\usepackage{epsfig}
\usepackage{graphicx}
\usepackage{amsmath,amssymb,amsfonts}
\usepackage{array}
\usepackage{url}
\usepackage{hyperref}
\usepackage{multirow}
\usepackage{float}
\usepackage{lineno}
\usepackage{xspace}
\usepackage[usenames,dvipsnames]{color}

\newcommand{\bef}{\begin{figure}}
\newcommand{\eef}{\end{figure}}
\newcommand{\bc}{\begin{center}}
\newcommand{\ec}{\end{center}}

\newcommand{\be}{\begin{equation}}
\newcommand{\ee}{\end{equation}}
\newcommand{\bea}{\begin{eqnarray}}
\newcommand{\eea}{\end{eqnarray}}

\begin{document}
\title{$J/\psi$ Production Dynamics: Event shape, Multiplicity and Rapidity dependence in Proton+Proton Collisions at LHC energies using PYTHIA8}
\author{Anisa Khatun}
\affiliation{Department of Physics, Aligarh Muslim University, Aligarh, 202002, India}
\author{Dhananjaya Thakur}
\author{Suman Deb}
\author{Raghunath Sahoo}
\email{Raghunath.Sahoo@cern.ch}
\affiliation{Discipline of Physics, School of Basic Sciences, Indian Institute of Technology Indore, Simrol, Indore 453552, INDIA}

\begin{abstract}
 High-multiplicity pp collisions at the Large Hadron Collider (LHC) energies have created special importance in view of the Underlying Event (UE) observables. The recent results of LHC, such as long range angular correlation, flow-like patterns, strangeness enhancement etc. in high multiplicity events are not yet completely understood. In the same direction, the understanding of multiplicity dependence of J/$\psi$ production is highly necessary. Transverse spherocity, which is an event shape variable, helps to investigate the particle production by isolating the hard and the soft components. In the present study, we have investigated the multiplicity dependence of J/$\psi$ production at mid-rapidity and forward rapidity through the transverse spherocity analysis and tried to understand the role of jets by separating the isotropic and jetty events from the minimum bias collisions. We have analyzed the J/$\psi$ production at the mid-rapidity and forward rapidities via dielectron and dimuon channels, respectively using 4C tuned PYTHIA8 event generator. The analysis has been performed in two different center-of-mass energies: $\sqrt{s}$ = 5.02 and 13 TeV, to see the energy dependence of jet contribution to the multiplicity dependence study of J/$\psi$ production. Furthermore, we have studied the production dynamics through the dependence of thermodynamic parameters on event multiplicity and transverse spherocity.  

\end{abstract}
\date{\today}
\maketitle

\section{Introduction}
\label{intro}
The observation of quark-gluon plasma (QGP) like effects in small systems (pp and p$-$A) continues to generate considerable interest in the scientific community, {\it e.g.} the discovery of collective-like phenomena~\cite{Li:2011mp,Khachatryan:2010gv}, strangeness enhancement~\cite{ALICE:2017jyt} etc. in high-multiplicity pp and p$-$A collisions. In this context, an important question arises, namely, do the QGP-like phenomena involve all the particles in the system, or it is just the effect of contributions from the processes like resonance decays, jets, underlying events (UE) etc. Therefore, the small systems need to be reinvestigated properly. To observe similar effects and in particular, the effect of UE 
on J/$\psi$ production, ALICE has performed the multiplicity dependence study of J/$\psi$ at mid- and forward rapidities~\cite{Abelev:2012rz,Adamova:2017uhu}. A faster than linear and approximately linear behavior has been observed at mid- and forward rapidities, respectively~\cite{Thakur:2019qau}. The faster than linear increase of J/$\psi$ yield with multiplicity questions the role of phenomena like collectivity, contribution of higher Fock states, color string percolation, color reconnection etc., in addition to the multipartonic interaction (MPI)~\cite{Weber:2017hhm,Sjostrand:2007gs,Ferreiro:2012fb,Kopeliovich:2013yfa,Thakur:2017kpv}. It has been speculated that different kind of trends for multiplicity dependence of J/$\psi$ at mid and forward rapidity might be due to auto-correlation and/or jet biases. Auto-correlation is an experimental effect which may arise when two observables are studied in the same phase space. Jet-bias is the possible effect of jets on the particle yields (here J/$\psi$) in the jet-rich environment.

To understand the production dynamics in a better way, a differential analysis involving tools to separate jetty from isotropic events thus becomes evident. The event shape analysis is a promising tool for analyzing the jet biases in high multiplicity pp events. The collision centrality, as well as the predominant reaction mechanisms at each centrality, can be inferred from the experimentally measured characteristics of the particle emission~\cite{Banfi:2010xy,Aad:2012fza,Ortiz:2017jho,Tripathy:2019blo,Khuntia:2018qox}. Therefore, event shapes measure the geometrical properties of the energy flow in QCD events. In the present work, we have performed an event shape analysis on the basis of transverse spherocity of mid-rapidity charged hadrons, applied to events generated with Pythia~8.2. This technique helps to isolate jetty-like (high-$p_{T}$ jets) and isotropic (low-$Q^{2}$ partonic scatterings) events~\cite{Cuautle:2015kra}, which helps in studying the physical observables separately in both the event types. Hence, spherocity can be used to study the possible soft and hard-QCD contributions to J/$\psi$ production from both kinds of events. In this work, we have performed the spherocity evolution of J/$\psi$ production with energy, multiplicity and rapidity. This study will help to understand different kinds of trends for J/$\psi$ as a function of multiplicity at mid and forward rapidities, particularly the jet biases in the J/$\psi$ production. Therefore, the present work aims to explore the following aspects,

\begin{itemize}
 \item Rapidity dependence of jet-bias to the multiplicity dependent J/$\psi$ production
 \item Energy dependence of jet-bias to the multiplicity dependent J/$\psi$ 
 \item Rapidity and energy dependence of production dynamics of J/$\psi$ through the dependencies of thermodynamic parameters on event shape and event multiplicity 
  \end{itemize}
  
  The analysis has been performed in simulated pp collisions at $\sqrt{s}$ = 5.02 and 13 TeV via the dielectron and dimuon decay channels of J/$\psi$ at mid and forward rapidities, in view of measurements by the ALICE experiment~\cite{Thakur:2019qau}. The systems created in pp collisions are not fully thermalized i.e. are away from thermal equilibrium. Therefore, this type of systems are better described by the Tsallis non-extensive statistics. As $\rm{J}/\psi$ are formed early in the collisions, the usage of Tsallis non-extensive statistics is  more appropriate to draw inference about system thermodynamics using $\rm{J}/\psi$. The use of Tsallis distribution in describing particle spectra is also motivated by the spectral shape of identified particles in hadronic collisions as observed in RHIC and LHC experiments \cite{star-prc75,phenix-prc83,alice1,alice2,alice3,Khuntia:2017ite}.
\newline
\newline
\newline
\textbf{Transverse spherocity:}
\newline
For an event, transverse spherocity is defined for a unit vector $\hat{n} (n_{T},0)$ which minimizes the ratio given by~\cite{Cuautle:2015kra,Abelev:2012sk,Tripathy:2019blo,Khuntia:2018qox},
\begin{eqnarray}
S_{0} = \frac{\pi^{2}}{4} \bigg(\frac{\Sigma_{i}~\vec p_{T_{i}}\times\hat{n}}{\Sigma_{i}~p_{T_{i}}}\bigg)^{2}.
\end{eqnarray}
By restricting it to the transverse plane, it avoids the biases from the boost along the beam direction. This variable ranges from {\it zero} for pencil-like events (di-jet events), to a maximum of {\it one} for circularly symmetric events (isotropic events), which corresponds to mainly hard events and soft events, respectively. A schematic picture showing the transverse spherocity distribution in a hadronic collision is shown in Fig.1.

\bef[ht]
 \bc
 \includegraphics[scale=0.4]{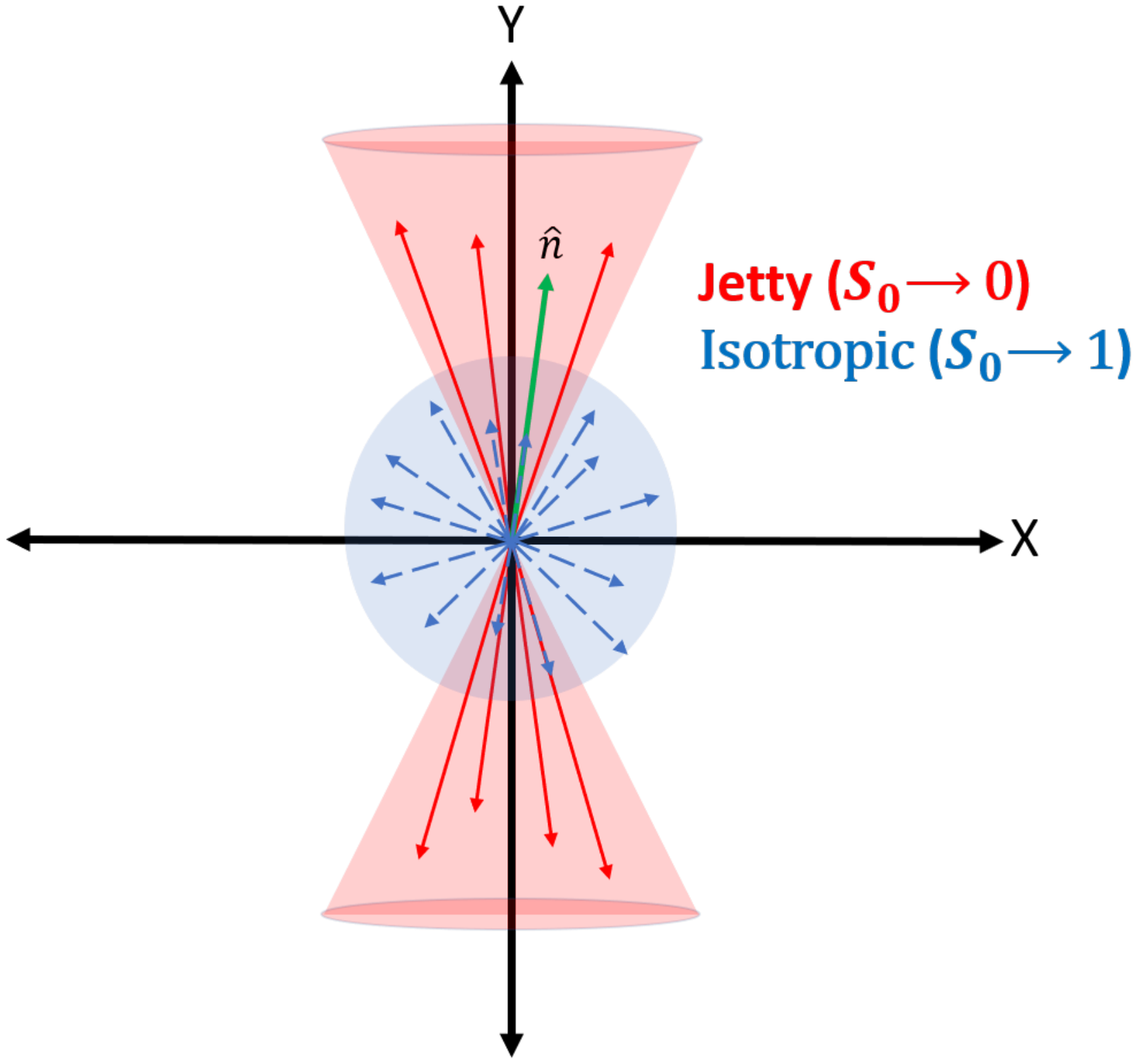}
\caption{ (Color online) A schematic picture showing transverse spherocity distribution of a hadronic collision.}
 \label{fig3}  
 \ec
 \eef

\textbf{Tsallis non-extensive statistics:}
\\
The transverse momentum $(p_{T})$ spectra of final state particles produced in high-energy collisions has been proposed to follow a thermalized Boltzmann type of distribution as given by  \cite{Hagedorn:1965st}

         \begin{eqnarray}
         \label{eq1}
         E\frac{d^3\sigma}{d^3p}& \simeq C \exp\left(-\frac{p_T}{T_{exp}}\right).
         \label{eq1}
         \end{eqnarray}
         
  To account for the high-$p_{\rm T}$ tail, a power-law in $p_{\rm T}$ is proposed \cite{CM,UA1}, which empirically accounts for the possible QCD contributions. Hagedorn proposed a combination of both the aspects, which describes the experimental data over a wide  $p_{\rm T}$ range  \cite{Hagedorn:1983wk} and is given by 
  
\begin{eqnarray}
  E\frac{d^3\sigma}{d^3p}& = &C\left( 1 + \frac{p_T}{p_0}\right)^{-n}
\nonumber\\
 & \longrightarrow&
  \left\{
 \begin{array}{l}
  \exp\left(-\frac{n p_T}{p_0}\right)\quad \, \, \, {\rm for}\ p_T \to 0, \smallskip\\
  \left(\frac{p_0}{p_T}\right)^{n}\qquad \qquad{\rm for}\ p_T \to \infty,
 \end{array}
 \right .
 \label{eq2}
\end{eqnarray}

where $C$, $p_0$, and $n$ are parameters.

 A thermodynamically consistent  non-extensive distribution called Tsallis distribution is given by~\cite{Cleymans:2011in,Thakur:2016boy}, 

 \begin{equation}
\label{eq3}
f(m_T) =  C_q \left[1+{(q-1)}{\frac{m_T}{T}}\right]^{-\frac{1}{q-1}} .
\end{equation}
  Here, $m_{\rm T}$ ($m_{\rm T} = \sqrt{p_{T}^{2} + m^{2}} )$ is the transverse mass and $q$ is called the non-extensive parameter-- a measure of degree of deviation from equilibrium. Eqs. \ref{eq2} and \ref{eq3} are related through the following transformations for large values of $p_{T}$:
  \begin{equation}
  n= \frac{1}{q-1}, ~\mathrm{and} ~~~~ p_0 = \frac{T}{q-1}.
  \label{eq4}
  \end{equation} 

\par
 In the limit $q \rightarrow 1$, one recovers the standard Boltzmann-Gibbs thermalized distribution (Eq. \ref{eq1}) from the Tsallis distribution. The transverse momentum spectra of J/$\psi$ are well described by a power-law function given by Eq.\ref{eq2}~\cite{Acharya:2017hjh,Acharya:2019lkw}. In this work, through its thermodynamical connection as given in Eq.~\ref{eq4}, we have performed the event shape as well as multiplicity dependence study of thermodynamic parameters of J/$\psi$ using the Tsallis non-extensive statistics.


\section{Event generation and Analysis methodology}
\label{eventgen}
PYTHIA8 is a Monte Carlo based pQCD inspired event generator. It is an improved version of PYTHIA6 which includes the implementation of MPI based scenario, where $2\rightarrow2$ hard sub-processes can produce heavy quarks like charm ($c$) and beauty ($b$). Elaborate description of PYTHIA8.2 physics processes can be found in Ref.~\cite{pythia8html}.~4C tuned PYTHIA8.2~\cite{Corke:2010yf} is used in the present study. The same tune has been used in our previous works to study the role of MPI, Colour Reconnection (CR) like UEs in $J/\psi$ production ~\cite{Thakur:2017kpv, Deb:2018qsl}. Varying impact parameter (MultipartonInteractions:bProfile=3~ switch of PYTHIA8) is included in the present study to allow all incoming partons to undergo hard and semi-hard interactions. MPI-based scheme of CR (ColourReconnection:mode(0)) of PYTHIA8.2 is used. More details on various models of CR included in PYTHIA8.2 and their performances with respect to experimental data can be found in Refs~\cite{Bierlich:2015rha,Abelev:2013bla}.
 
 In the present work, inelastic, non-diffractive component of the total cross section for all hard QCD processes (HardQCD:all=on) are simulated, which includes the production of heavy quarks. A $p_{T}$ cut-off of $p_{T} =$ 0.5 GeV/c (using PhaseSpace:pTHatMinDiverge) is imposed to avoid the divergences of QCD processes in the limit $p_{T}\rightarrow0$. We have specifically decayed $J/\psi \rightarrow e^{+} + e^{-}$  at the mid-rapidity ($|y| < 0.9$) and $J/\psi \rightarrow \mu^{+} + \mu^{-}$ at the forward rapidity ($2.5 < y < 4.0$). The measurement of J$/\psi$ yields are done through invariant mass reconstruction by using the above external decay modes to follow the ALICE experimental results and make an explicit comparison with them. Figure~\ref{fig2} and Fig.~\ref{fig3} show the comparison of  experimental data of ALICE with PYTHIA8. From these figures, it can be seen that PYTHIA8 well reproduces the ALICE measurements.
 For further studies of event shape analysis, we have used these settings in PYTHIA8 because of the agreement with experimental results.

\bef[ht]
 \bc
\includegraphics[scale=0.42]{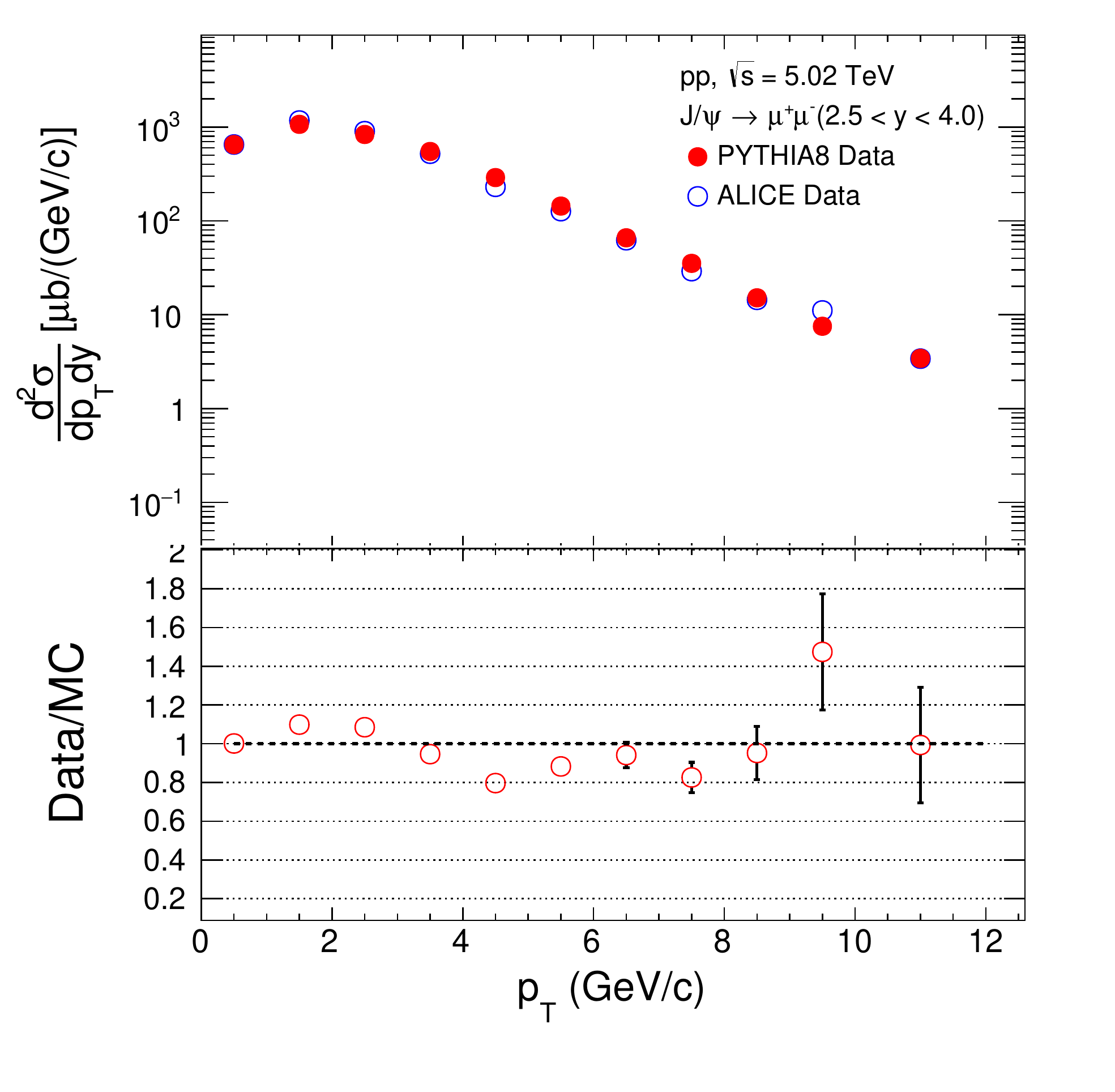}
\caption{ (Color online) Top panel shows the comparison of ALICE data~\cite{Acharya:2017hjh} and PYTHIA8 of $J/\psi$ production cross-section as a function of transverse momentum for pp collisions at $\sqrt{s}$ = 5.02 TeV. The open blue circles are ALICE data and solid red circles represent PYTHIA8 results. The quadratic sum of statistical and systematic uncertainties of ALICE data are presented in a single error bar.  Bottom panel shows the ratio between ALICE data and PYTHIA8, and the error bars are estimated using standard error propagation formula, where quadratic sum of statistical and systematic uncertainties of ALICE data are taken into account. A constant multiplier of 0.032 has been applied to PYTHIA8 data for a matching of spectral shape with the corresponding experimental data.}
 \label{fig2}  
 \ec
 \eef
 
 \bef[ht]
 \bc
\includegraphics[scale=0.42]{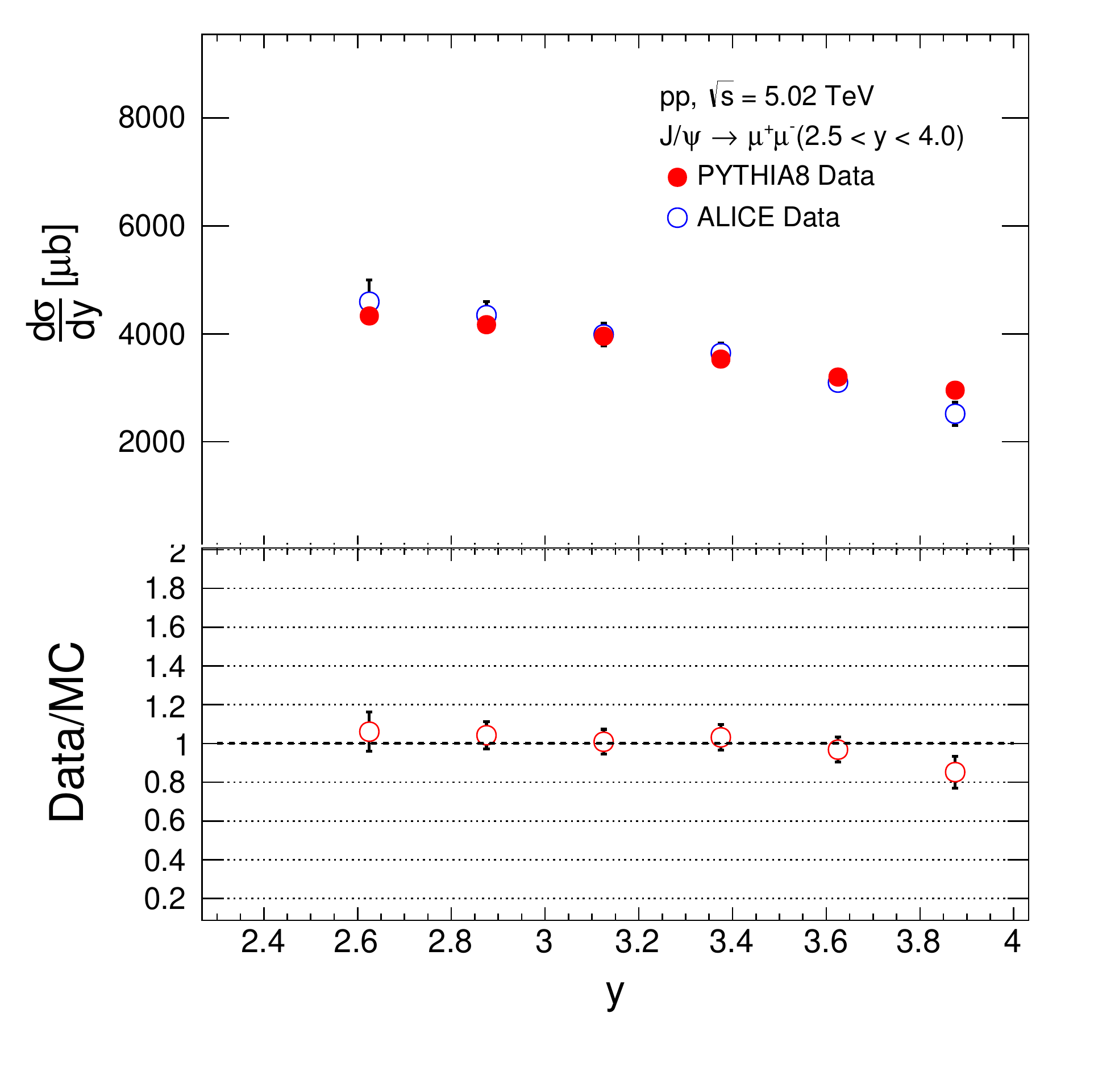}
\caption{ (Color online) Top panel shows the comparison of ALICE data~\cite{Acharya:2017hjh}  and PYTHIA8 for $J/\psi$ production cross-section as a function of rapidity for pp collisions at $\sqrt{s}$ = 5.02 TeV. The open blue circles are ALICE data and solid red circles represent the PYTHIA8 results. The quadratic sum of statistical and systematic uncertainties of ALICE data are presented in a single error bar.  Bottom panel shows the ratio between ALICE data and PYTHIA8, and the error bars are estimated using standard error propagation formula, where quadratic sum of statistical and systematic uncertainties of ALICE data are taken into account. A constant multiplier of 0.18 has been applied to PYTHIA8 data for a matching of spectral shape with the corresponding experimental data.}
 \label{fig3}  
 \ec
 \eef

We have generated (6.1$\times10^{8}$, 1.22$\times10^{8}$) and  (9.6$\times10^{8}$, 1.14$\times10^{8}$) events for pp collisions at $\sqrt{s}$ = 5.02 and 13 TeV at (forward, mid)-rapidities, respectively. The charged particle multiplicity, $N_{ch}$ is measured at the mid-rapidity ($|\eta|$ $<$ 1.0).
The spherocity distributions are selected in the same pseudo-rapidity range with a minimum constraint of 5 charged particles with $p_{T}>$ 0.15 GeV/c. The jetty events are the lowest 20$\%$ and the isotropic events are highest  20$\%$ of the total events and these correspond to the ranges: $0 <S_0<0.37$ and $0.71< S_0 <1$, respectively. Figure.~\ref{fig4} represents the spherocity distribution in different multiplicity intervals for pp collisions at $\sqrt{s}$ = 5.02 TeV. Here, it is observed that high-multiplicity events are more towards isotropic in nature. The process of isotropization in a multiparticle final state happens through multiple interactions between the quanta of the system~\cite{Cuautle:2015kra}. When the final state multiplicity in an event is higher, the probability of the event becoming isotropic is also higher.

\bef[ht]
 \bc
 \includegraphics[scale=0.42]{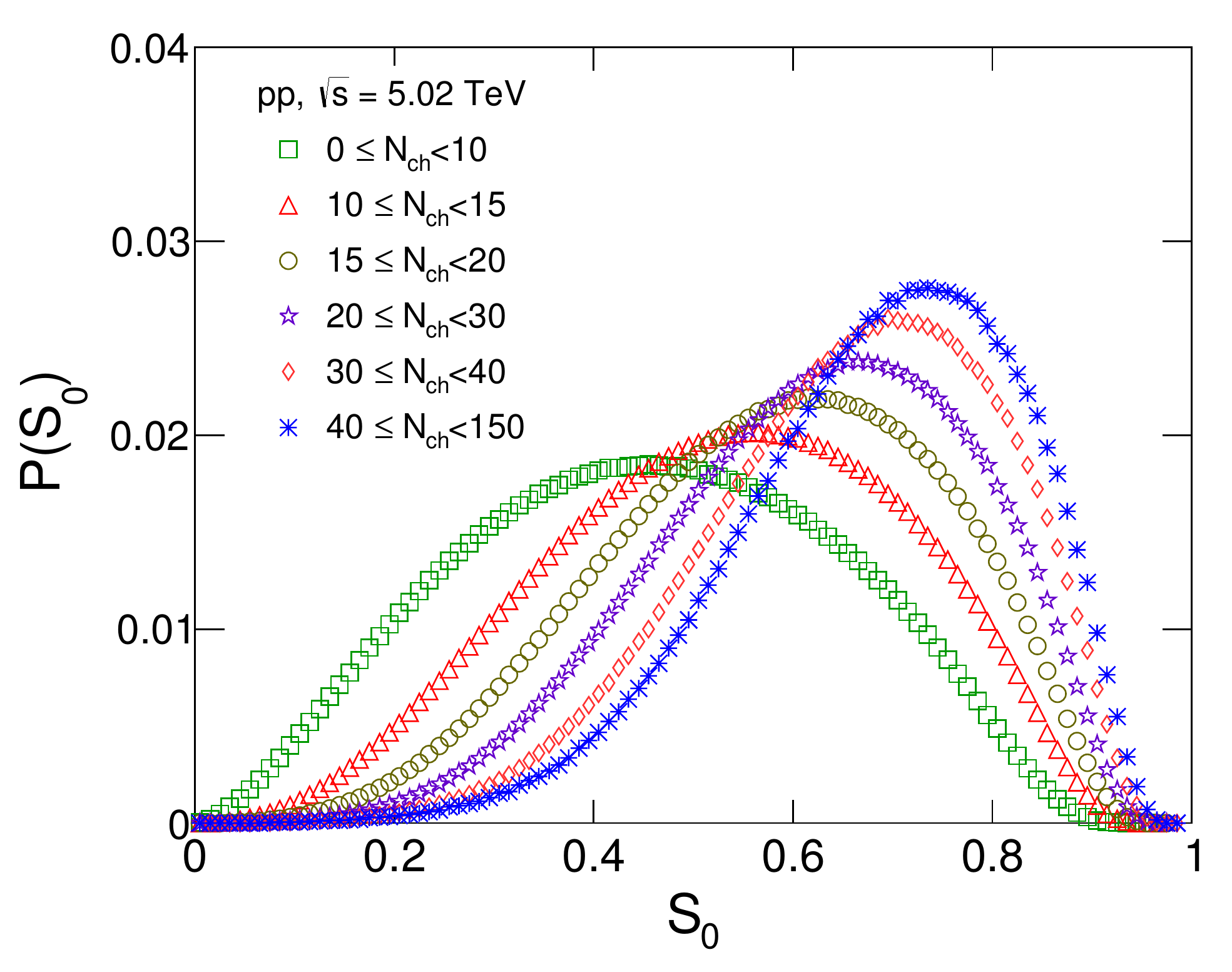}
\caption{ (Color online) The spherocity distribution of minimum bias events as a function of charged particle multiplicity in pp collisions at $\sqrt{s}$ = 5.02 using PYTHIA8.}
 \label{fig4}  
 \ec
 \eef

\section{Results and Discussion}
 \label{result}
 
 At the LHC energies, MPI is the process which occurs at a substantial rate in hadronic collisions and is a very important ingredient in explaining the multiplicity dependence of various observables. MPI incorporated in PYTHIA8, is able to describe many of the experimentally observed features, such as multiplicity distribution and underlying events; light-flavour and heavy-flavor production; and together with color-reconnection, it is able to reproduce flow-like patterns in pp collisions \cite{Ortiz:2017jaz, ALICE:2017pcy,Acharya:2018orn,Adam:2015qaa,Colamaria:2015wxs,Ortiz:2013yxa}. Models containing MPI well describe the multiplicity dependence of J/$\psi$ production and thus reveal that MPI is an important mechanism behind the production of J/$\psi$~\cite{Weber:2017hhm,Thakur:2017kpv}. In a given multiplicity interval, there are events originating from different number of MPIs, which make them different in nature. As discussed in the previous section, using the transverse spherocity, we can classify the events based on their jet content. Quarkonium production in the parton shower, which is able to explain the lack of observed polarization, predicts that J/$\psi$ mesons are rarely produced in isolation in hadronic collisions~\cite{Baumgart:2014upa,Aaij:2017fak}. Further, the production cross-section of J/$\psi$ at mid-rapidity is higher as compared to forward rapidity~\cite{Acharya:2019lkw}. This indicates the difference in jet contribution at mid and forward rapidity J/$\psi$ production. In this contribution, we have tried to investigate the production of J/$\psi$ in high-multiplicity pp collisions by analysing their $p_{\rm{T}}$ spectra in different jet environments.
  
 \subsection{Event shape dependence of J/$\psi$ production at mid and forward rapidities}
  \label{muldep}
  An event shape study is carried out using transverse spherocity in different charged-particle multiplicity classes (given in Table \ref{table1}). Figure~\ref{5tevpT_iso_jetty} shows the transverse momentum spectra of J/$\psi$ at the mid-rapidity (left panel) and at the forward rapidity (right panel) for integrated multiplicity (minimum bias) at $\sqrt{s}$ = 5.02 TeV for different spherocity classes. The lower pannel of the same figure represents the ratio of $p_{\rm{T}}$-spectra for isotropic and jetty events with respect to spherocity integrated events \textbf{($0 < S_0< 1.0$)}. It can be seen that the lower $p_T$ region is dominated by isotropic events over the jetty events. However, this scenario reverses as we move towards higher $p_T$. At a particular point, termed as  `crossing point', the jetty events dominate over the isotropic events. Therefore, the study of the `crossing point' is of great interest as far as a feasible boundary for dominance of the event type and hence the associated particle production mechanisms are concerned.  
 
\bef[ht]
\bc
\includegraphics[scale=0.45]{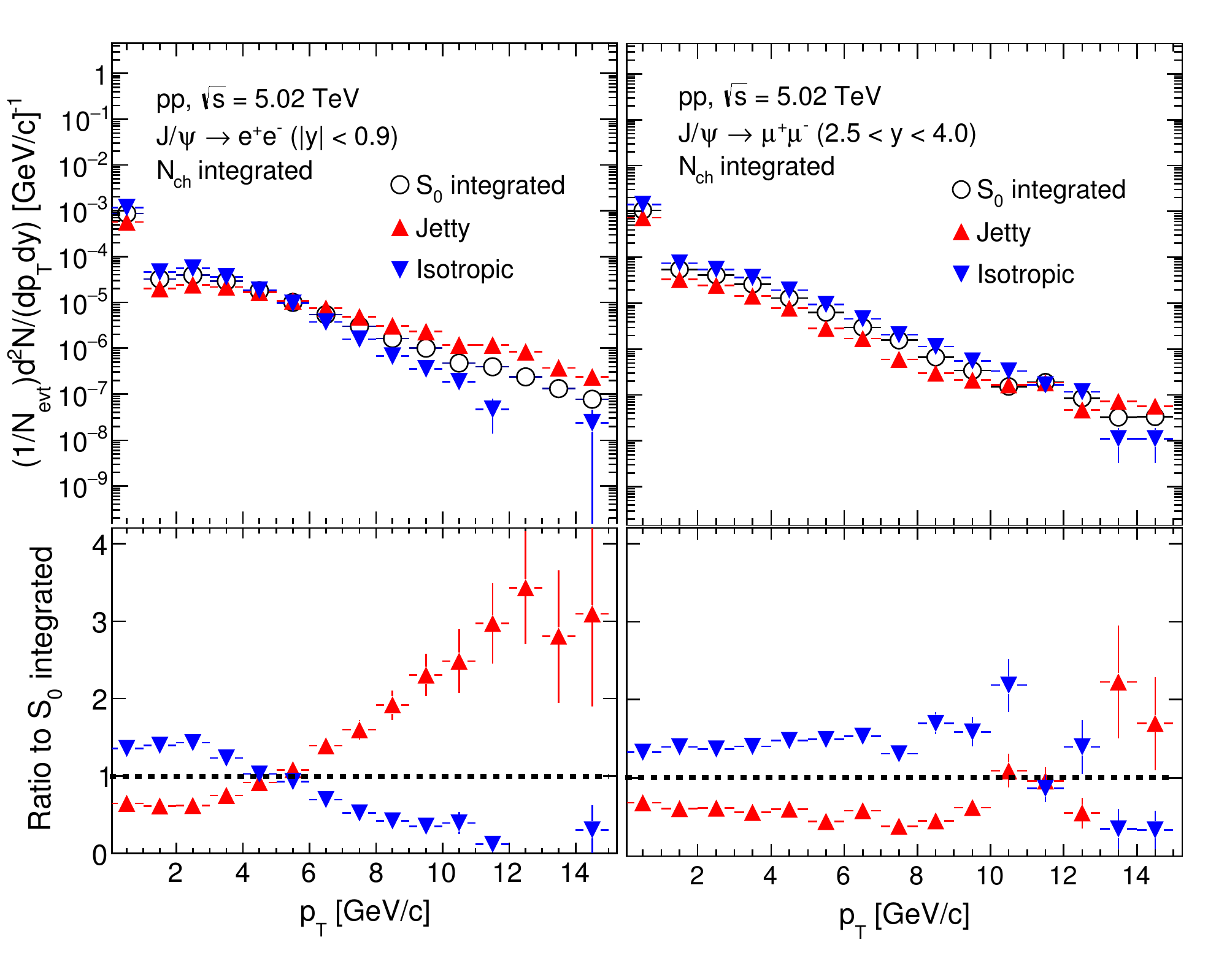}
\caption  {(Color online)  $p_{\rm{T}}$-spectra of J/$\psi$ produced using PYTHIA8 for  minimum  bias  pp  collisions  as  a function of spherocity for the mid- (left panel) and forward rapidities (right panel), respectively. The lower panel of the respective figures show the ratio of $p_{\rm{T}}$-spectra for isotropic and jetty events w.r.t $S_{\rm{0}}$ integrated events. The blue inverted triangles are for isotropic events, red triangles are jetty events, and open circles represent $S_{\rm{0}}$ integrated events.}
 \label{5tevpT_iso_jetty}  
\ec
 \eef
 
From the previous studies of the light-flavor sector, it has been observed that the `crossing point' largely depends on the multiplicity and the isotropic events are populated over jetty events as we move from low-multiplicity to high-multiplicities~\cite{Bencedi:2018ctm, Acharya:2019idg,Khuntia:2018qox}. The observations indicate that the QGP-like effects seen in high-multiplicity pp collisions are not because of jet-bias effects rather may be due to a possible system formation, which should be explored. In the present work, similar studies reveal that the speculation is valid in case of J$/\psi$ as well. From the current double differential study of J/$\psi$ {\it i.e.} J/$\psi$ as a function of spherocity and multiplicity, we have found that isotropic events are dominant at high-multiplicities. Motivated from the recent results of multiplicity dependence of J/$\psi$ production by ALICE~\cite{Thakur:2019qau}, we extend the analysis to look into the double differential study of rapidity dependence of J$/\psi$ production at mid- and forward rapidities. Figure~\ref{fig6} represents the  `crossing point' in different ranges of multiplicity, rapidity, and center-of-mass energy. Table~\ref{table1} presents the results, which are plotted in Fig~\ref{fig6}. This signifies how the contribution of jets to the production of $\rm{J}/\psi$ vary with rapidity and collision energy in different multiplicity environments. From Fig~\ref{fig6} as well as Table~\ref{table1}, a comparison of forward versus mid-rapidity reveals the shift of the `crossing point' towards lower-$p_T$ in case of mid-rapidity spherocity dependent J/$\psi$  production, indicating the jet dominant contribution of $\rm{J}/\psi$ production at the mid-rapidity. This means the contribution of jets in J/$\psi$ production is higher at the mid-rapidity.

\bef[ht]
 \bc
 \includegraphics[scale=0.45]{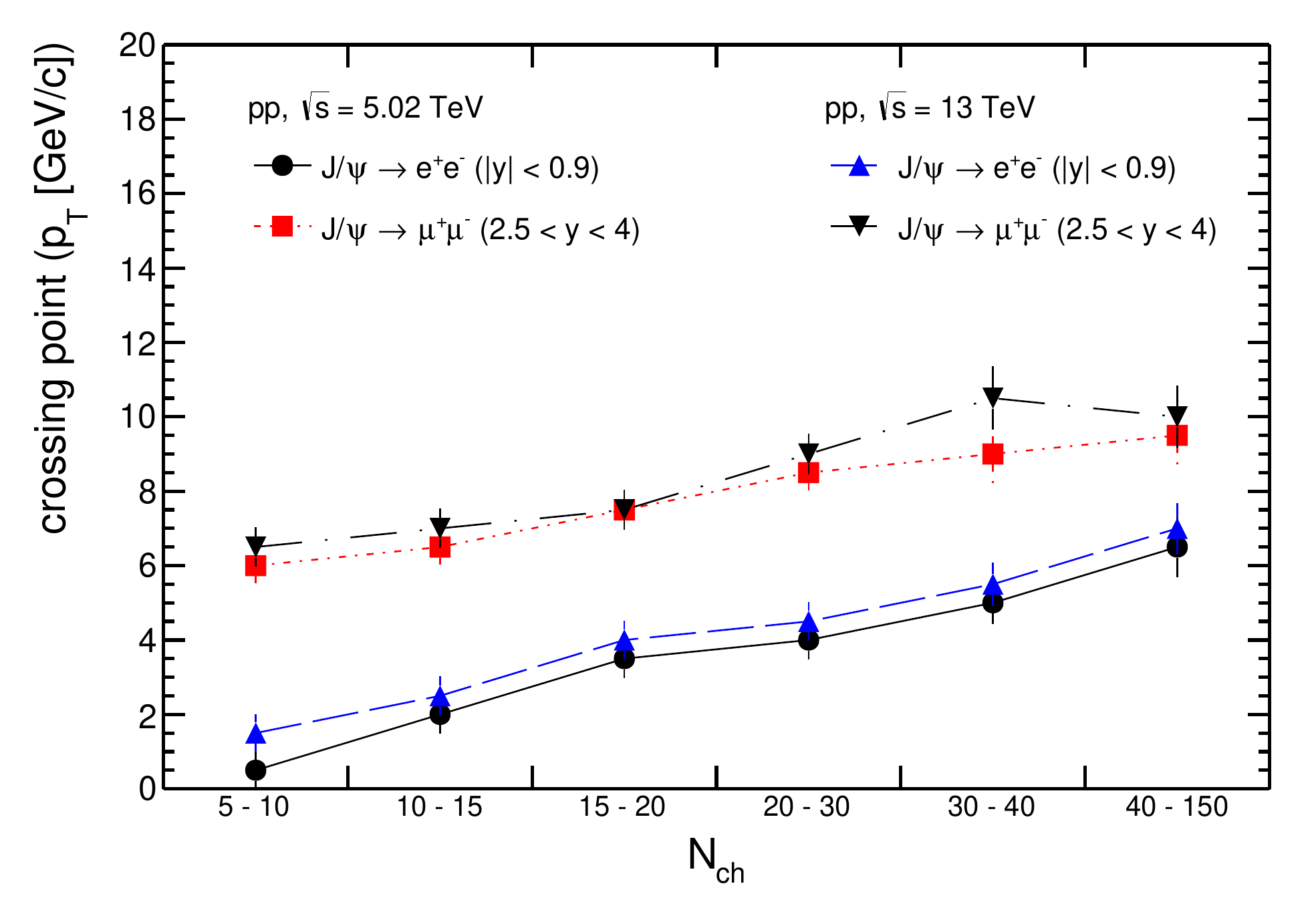}
\caption {(Color online)  The rapidity and energy dependence of `crossing point' to the $J/\rm{\psi}$ production as a function of event multiplicity.}
\label{fig6} 
 \ec
 \eef

\begin{table}[ht]
\centering
\caption{The crossing point ($p_T$ in GeV/c) of jetty and isotropic events at forward and mid-rapidity for pp collisions at $\sqrt{s}$ = 5.02 and 13 TeV.}
\begin{tabular}{|c|c|c|c|c|}
\hline
Multiplicity & \multicolumn{2}{c|}{$\sqrt{s}$ =  5.02 TeV} & \multicolumn{2}{c|}{$\sqrt{s}$ =  13 TeV} \\ \hline
$N_{ch}$ bin & $|y|<$ 0.9 & 2.5$<y<$4.& $|y|<$ 0.9 & 2.5$<y<$4.    \\ \hline
5-10    	& 0.50$\pm$ 0.50	& 6.00$\pm$0.57 	& 1.50$\pm$0.50   	& 6.50$\pm$0.53  \\ \hline
10-15   	& 2.00$\pm$ 0.52  	& 6.50$\pm$0.57 	& 2.50$\pm$0.52  	& 7.00$\pm$0.53 \\ \hline
15-20   	& 3.50$\pm$ 0.52	& 7.50$\pm$0.56 	& 4.00$\pm$0.52  	& 7.50$\pm$0.54 \\ \hline
20-30  	& 4.00$\pm$ 0.52	& 8.50$\pm$0.69 	& 4.50$\pm$0.53  	& 9.00$\pm$0.55 \\ \hline
30-40   	& 5.00$\pm$ 0.57 	& 9.00$\pm$0.80	& 5.50$\pm$0.58  	& 10.50$\pm$0.85 \\ \hline
40-150 	& 6.50$\pm$ 0.82 	& 9.50$\pm$0.90 	& 7.00$\pm$0.68  	& 10.00$\pm$0.69 \\ 
\hline
\end{tabular}
\label{table1}
\end{table}

Furthermore, the dominance of jettiness in low-multiplicity events and isotropiness in high-multiplicity events in $\rm{J}/\psi$ production reveals apparent reduction and softening of the jet yields at high$-N_{\rm{ch}}$~\cite{Chatrchyan:2013ala,Khuntia:2018qox}. In QCD inspired models like PYTHIA8, the prime mechanism to produce high-multiplicity pp events is related to partonic interactions with large momentum transfer. Therefore, the reduction in jet contribution in the $\rm{J}/\psi$ production may indicate a reduced production of back-to-back jets~\cite{Ortiz:2017jho}. Recently, in high-multiplicity pp collisions, away-side ridge in two-particle correlation has been observed, which is an indication of the presence of collectivity effects in the system~\cite{Khachatryan:2016txc}.
Our current observations go in the same direction with the observation of populated isotropic events at high-multiplicities. Hence, the contribution of jets to the production of $\rm{J}/\psi$ has very little but significant effect at high-multiplicities. Also, jet contribution in $\rm{J}/\psi$ production is significantly larger at mid-rapidity compared to forward rapidity. High multiplicities are rich with isotropic events which may give rise to collective-like effects, but the difference in jet bias at both the rapidities might produce different multiplicity dependent trends in $\rm{J}/\psi$ production. Furthermore, the dependence of the crossing point values with $N_{\rm{ch}}$ seems to increase moderately, appearing like a plateau for $N_{\rm{ch}} >$ 30 within uncertainty at forward-rapidity, whereas it goes on increasing for mid-rapidity. It shows that jet-contribution to $\rm{J}/\psi$ production at forward-rapidity tends to saturate beyond $N_{\rm{ch}} \approx$ 30. 

 \subsection{Energy dependence of J/$\psi$ production in different event shapes}
 \label{energydep}
 To investigate the possible dependence of jet effects on center-of-mass energy, we have performed the event activity dependence of double differential studies at $\sqrt{s} =$ 5.02 and 13 TeV. The Fig.~\ref{fig6} and Table~\ref{table1} represent the energy dependent contribution of jetty and isotropic events to the $\rm{J}/\psi$ production. The observation shows that although there is a large rapidity dependence of jet contribution to $\rm{J}/\psi$ production, it has very weak dependence on center-of-mass energy. From the Table~\ref{table1}, at a particular center-of-mass energy and rapidity, the value of the `crossing point' is the same within the uncertainty. This observation goes in line with the multiplicity dependence of $\rm{J}/\psi$ production as reported by ALICE experiment~\cite{Thakur:2019qau}, where the $\rm{J}/\psi$ production as a function of multiplicity is found to be independent of $\sqrt{s}$ at a given rapidity. Therefore, the `crossing point' depends on rapidity and is nearly independent of center-of-mass energy. This supports the recent observation by ALICE, where no dependence of $\rm{J}/\psi$ production on $\sqrt{s}$ has been observed~\cite{Thakur:2019qau}. 

\subsection{Event shape dependence of system thermodynamics}
\label{non_extensive}

As discussed in the previous sections, the production of $\rm{J}/\psi$ in jetty events is different from isotropic ones. While the former involves high-$p_{\rm{T}}$ phenomena, the latter is dominated by soft-physics. Tsallis non-extensive statistics is an appropriate model to describe all the three aspects of the event types, namely, $S_{0}$ integrated, jetty and isotropic events. The non-extensive parameter ($q$) gives the information about the degree of deviation of a system from thermodynamic equilibrium. Tsallis distribution function contains another parameter, ``T", called Tsallis temperature, which gives the information about the temperature of the system~\cite{Marques:2015mwa}. To study the event shape and multiplicity dependence of ``q" and ``T", we have fitted Tsallis distribution function to the transverse momentum spectra in spherocity and multiplicity classes for pp collisions at $\sqrt{s}=$ 5.02 and 13 TeV.  

\bef[ht]
\bc
\includegraphics[scale=0.48]{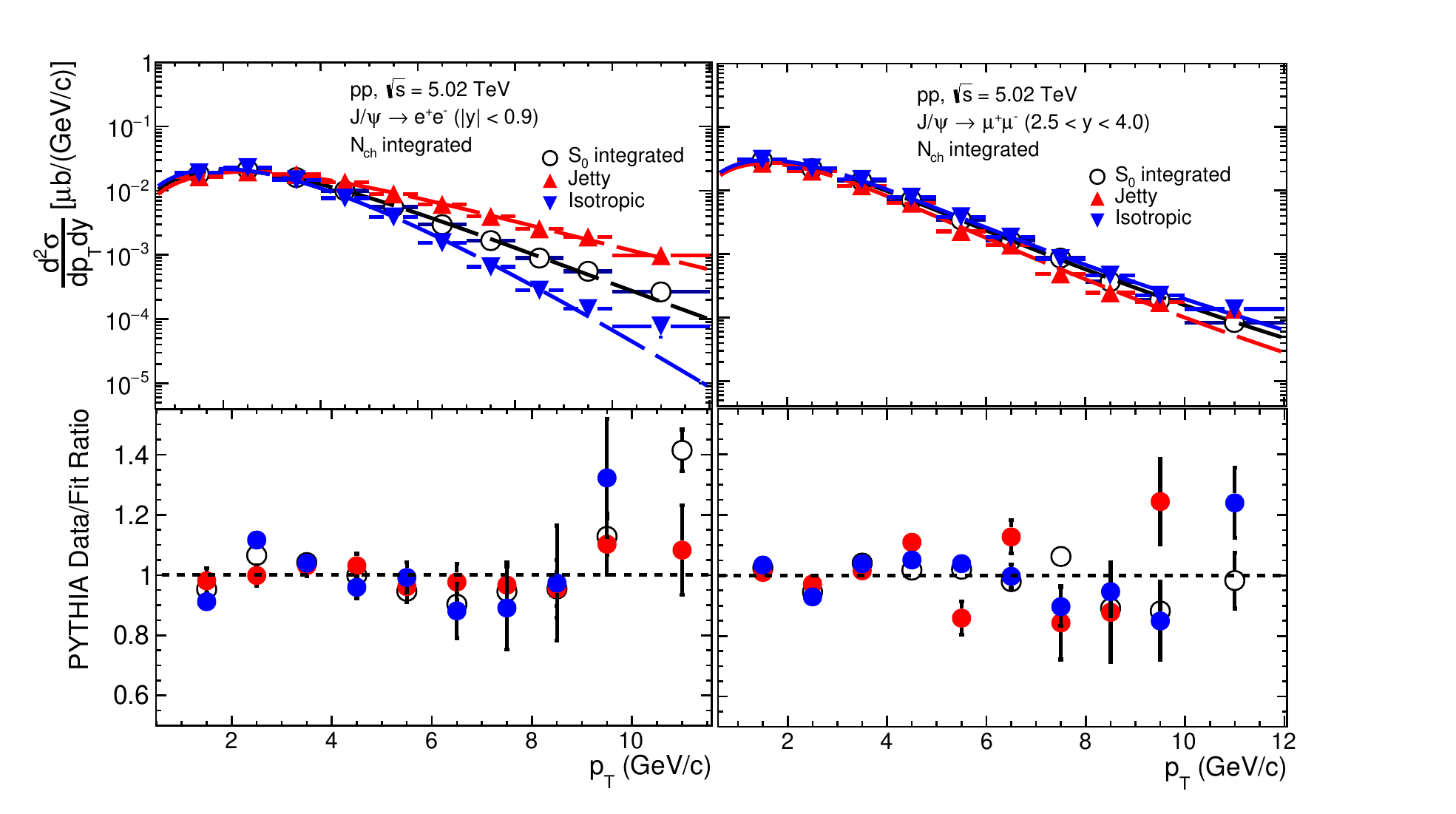}
\caption {(Color online) The $J/\psi$ production cross-section as a function of $p_T$ for minimum bias collisions in different event samples (jetty and isotropic) at mid (left panel) and forward rapidity (right panel). The spectra are described by Tsallis distribution function,  given by Eqn. \ref{eq3}. The bottom panels show the PYTHIA data to fit ratios for the respective rapidities.}
 \label{fig7}  
 \ec
 \eef

\begin{figure*} [ht]
\begin{center}
\includegraphics[width=18.2pc]{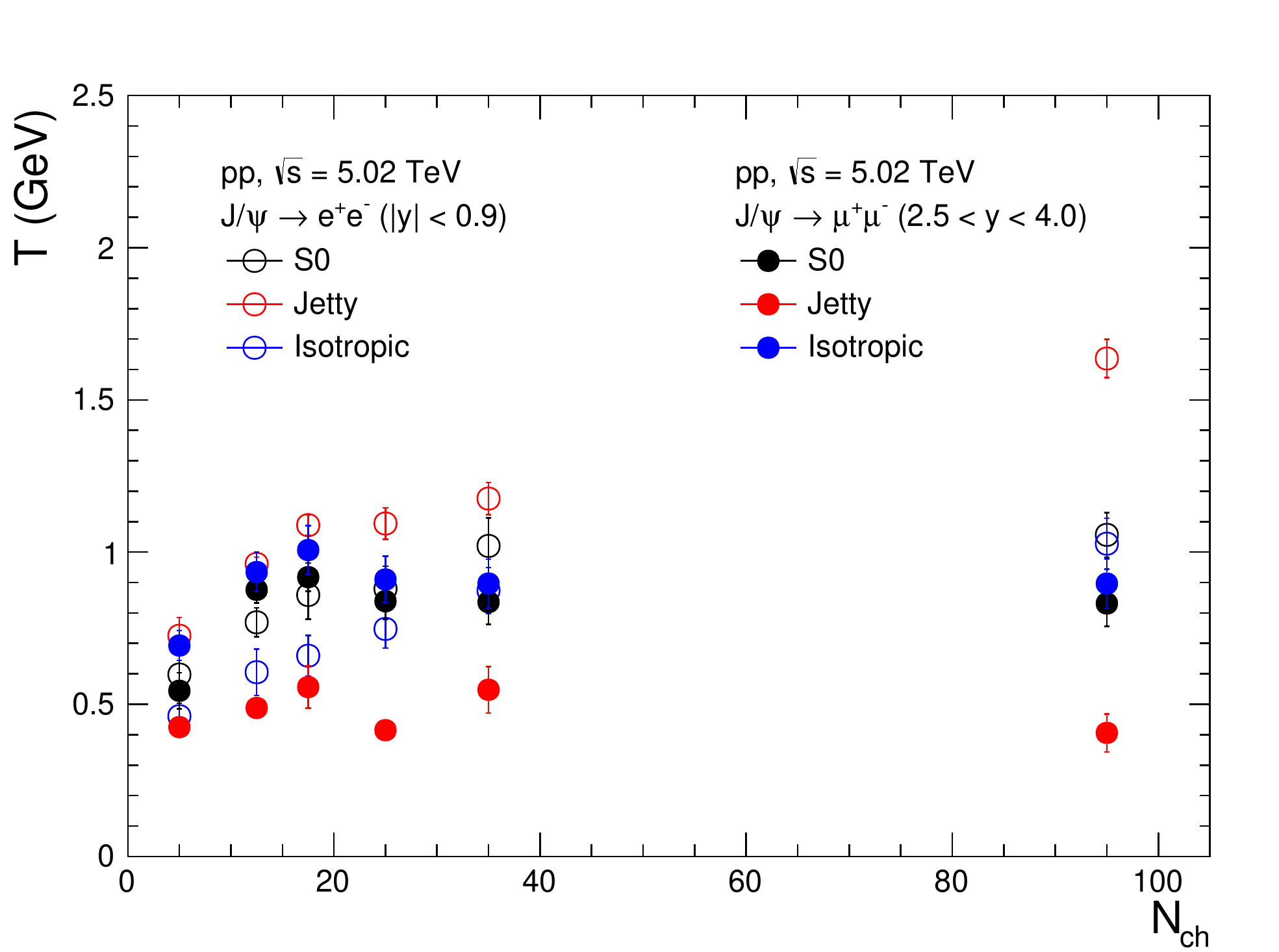}
\includegraphics[width=19.2pc]{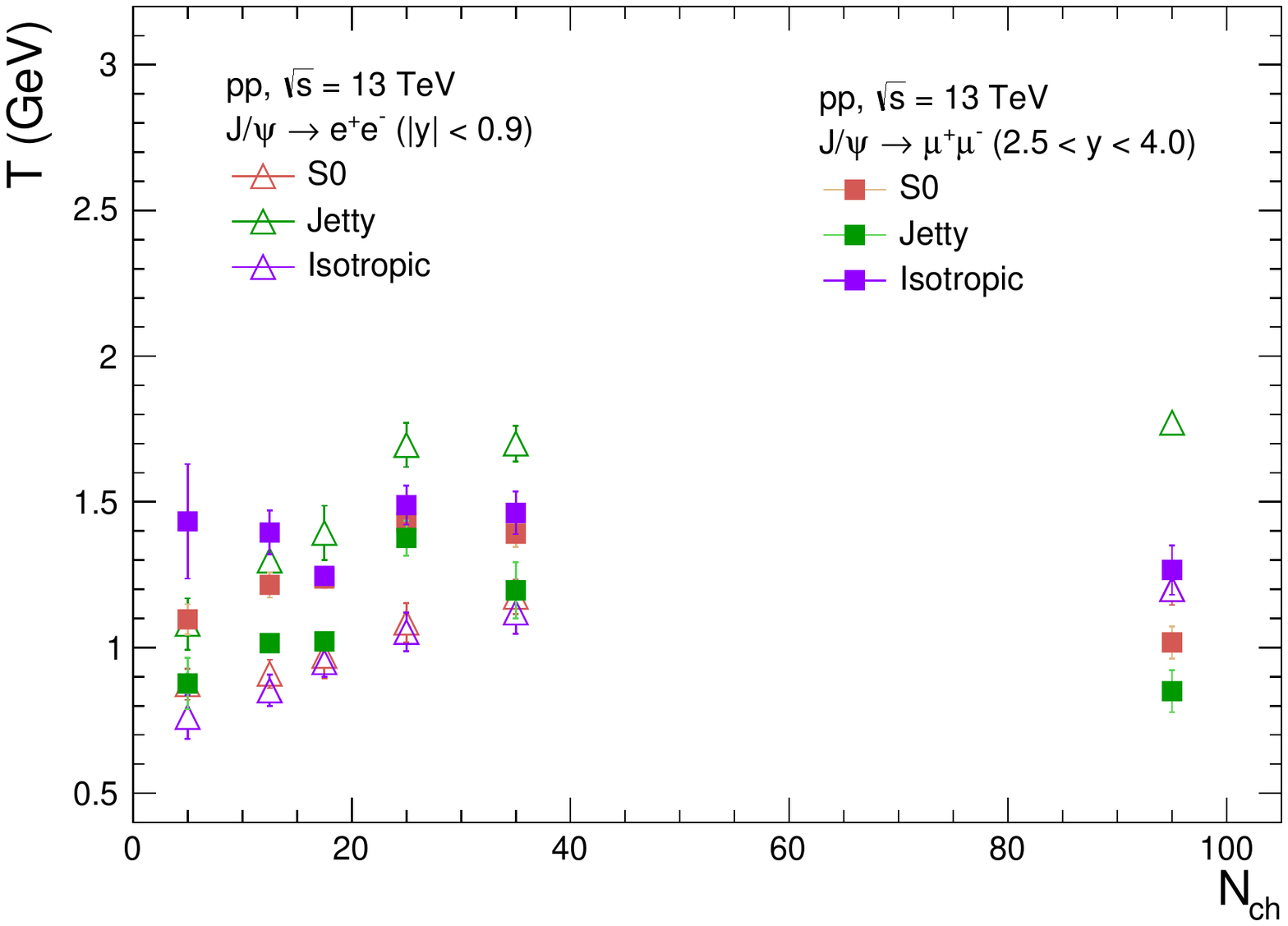}
\includegraphics[width=18.2pc]{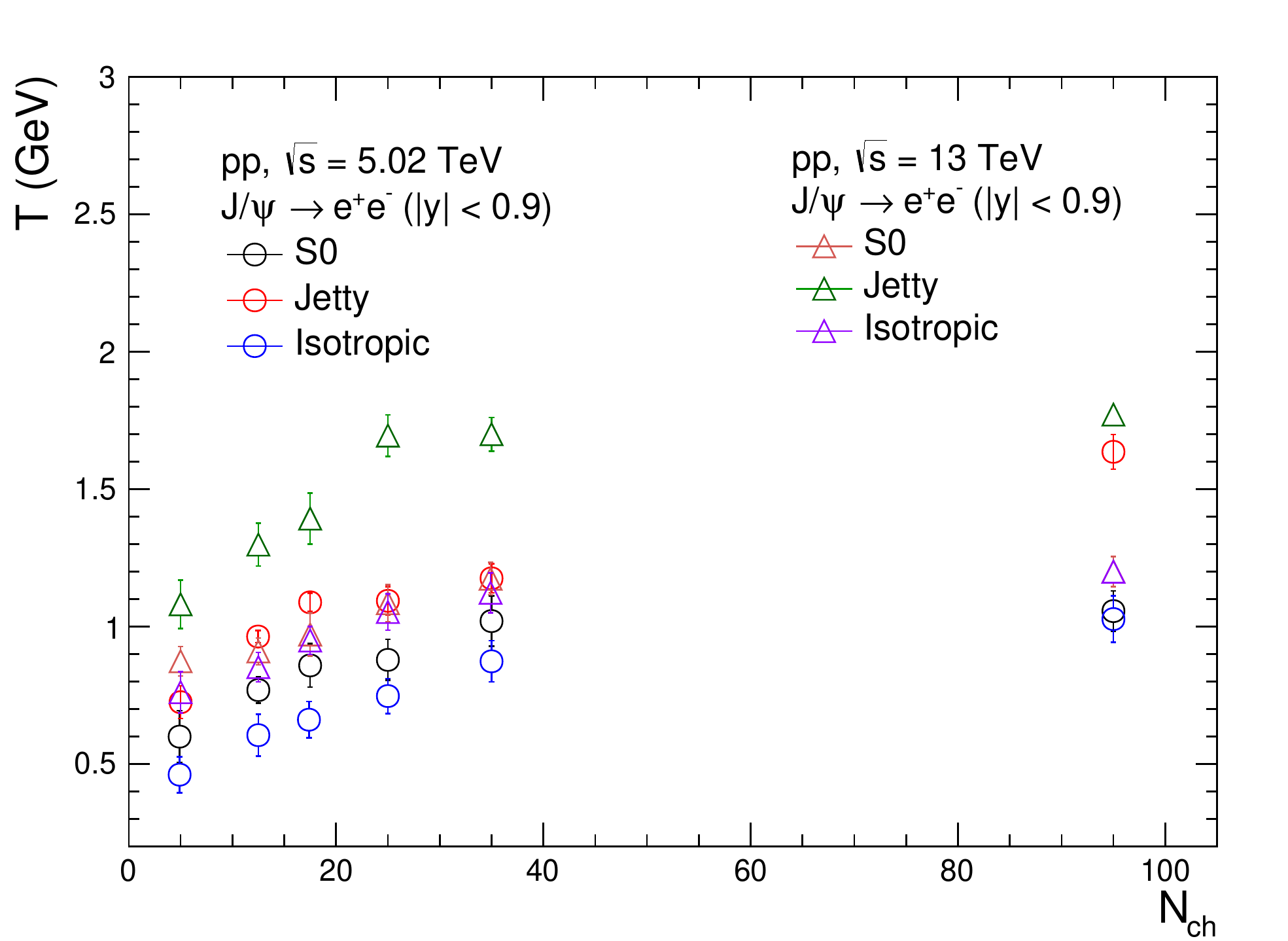}
\includegraphics[width=19.2pc]{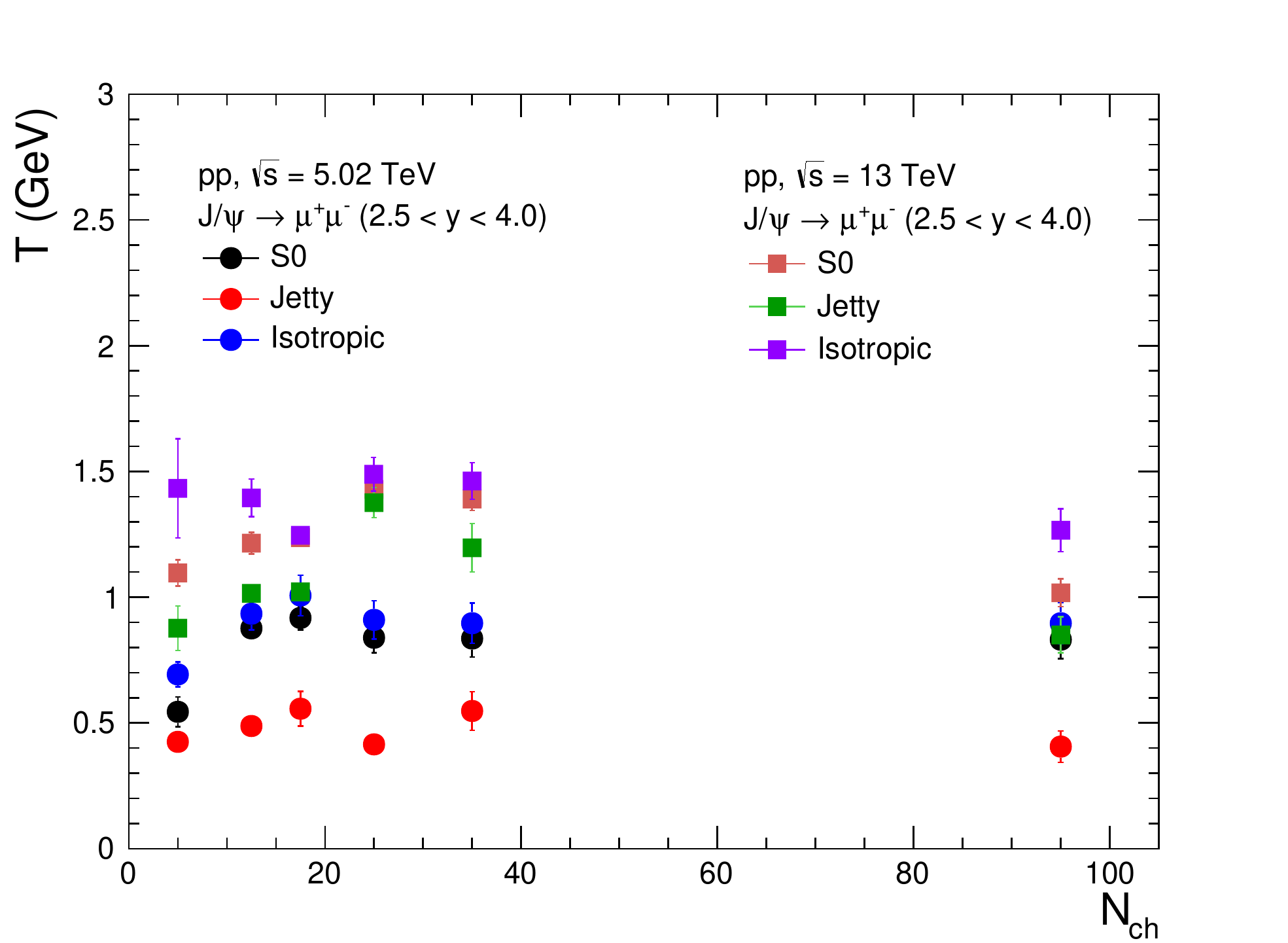}
\caption {(Color online) Multiplicity, rapidity and energy dependence of Tsallis temperature for $J/\psi$ at mid and forward rapidity for pp collisions at $\sqrt{s}$ = 5.02 and 13 TeV. The upper two panels represent the rapidity dependence and the lower two panels show energy dependent behavior.}
\label{fig:Temp}
\end{center}
\end{figure*}

\begin{figure*} [ht]
\begin{center}
\includegraphics[width=18pc]{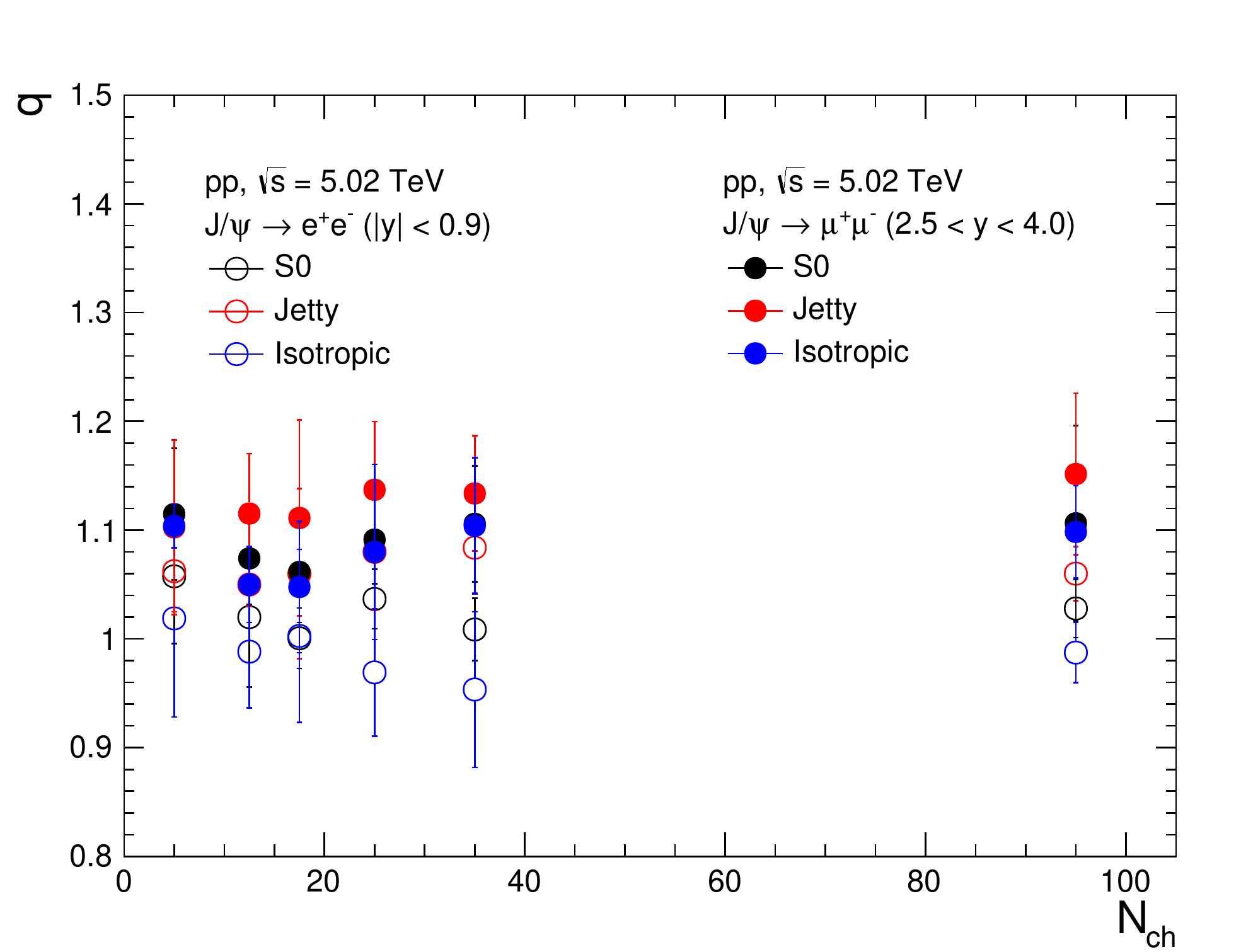}
\includegraphics[width=18.8pc]{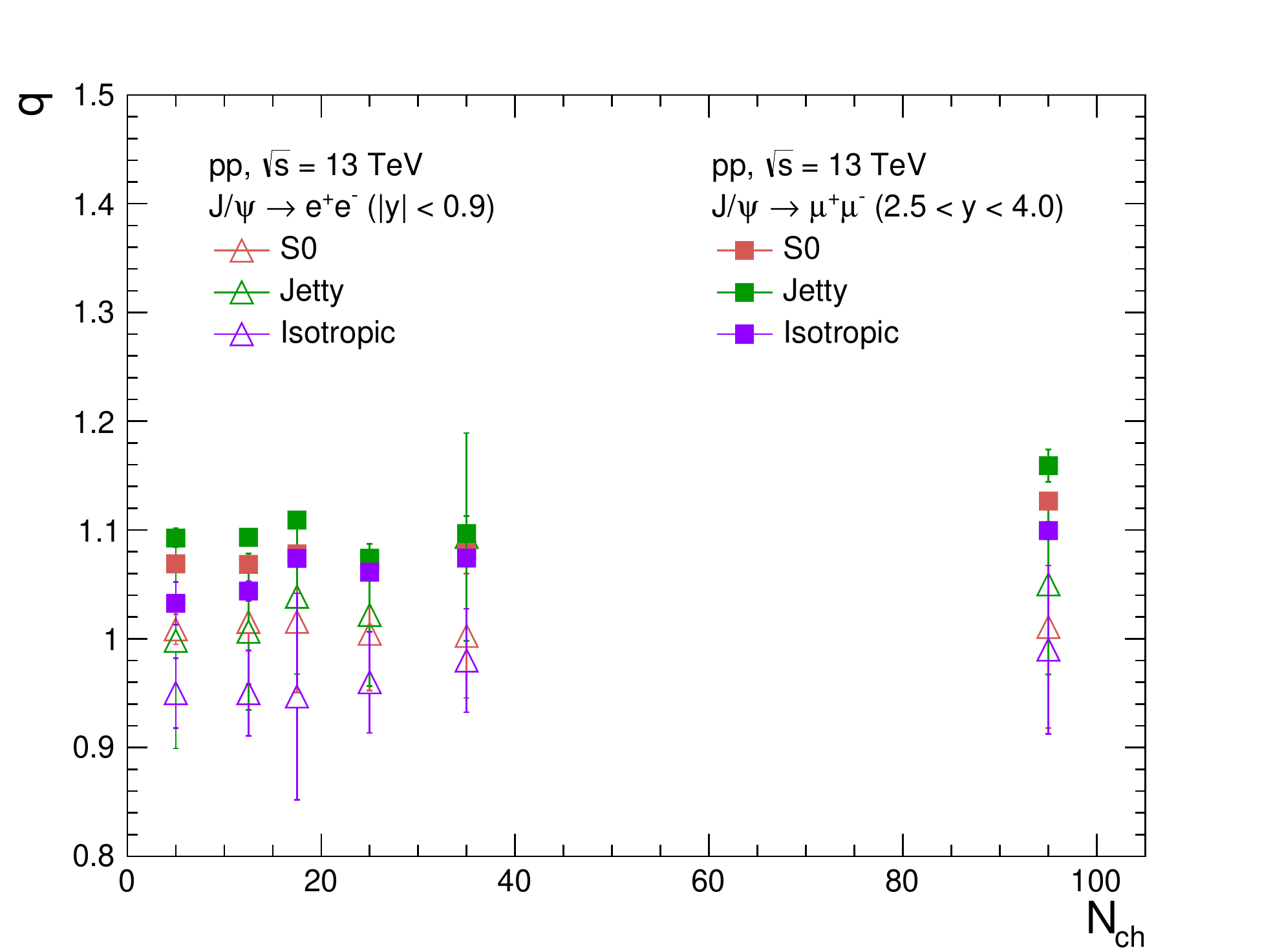}
\includegraphics[width=18pc]{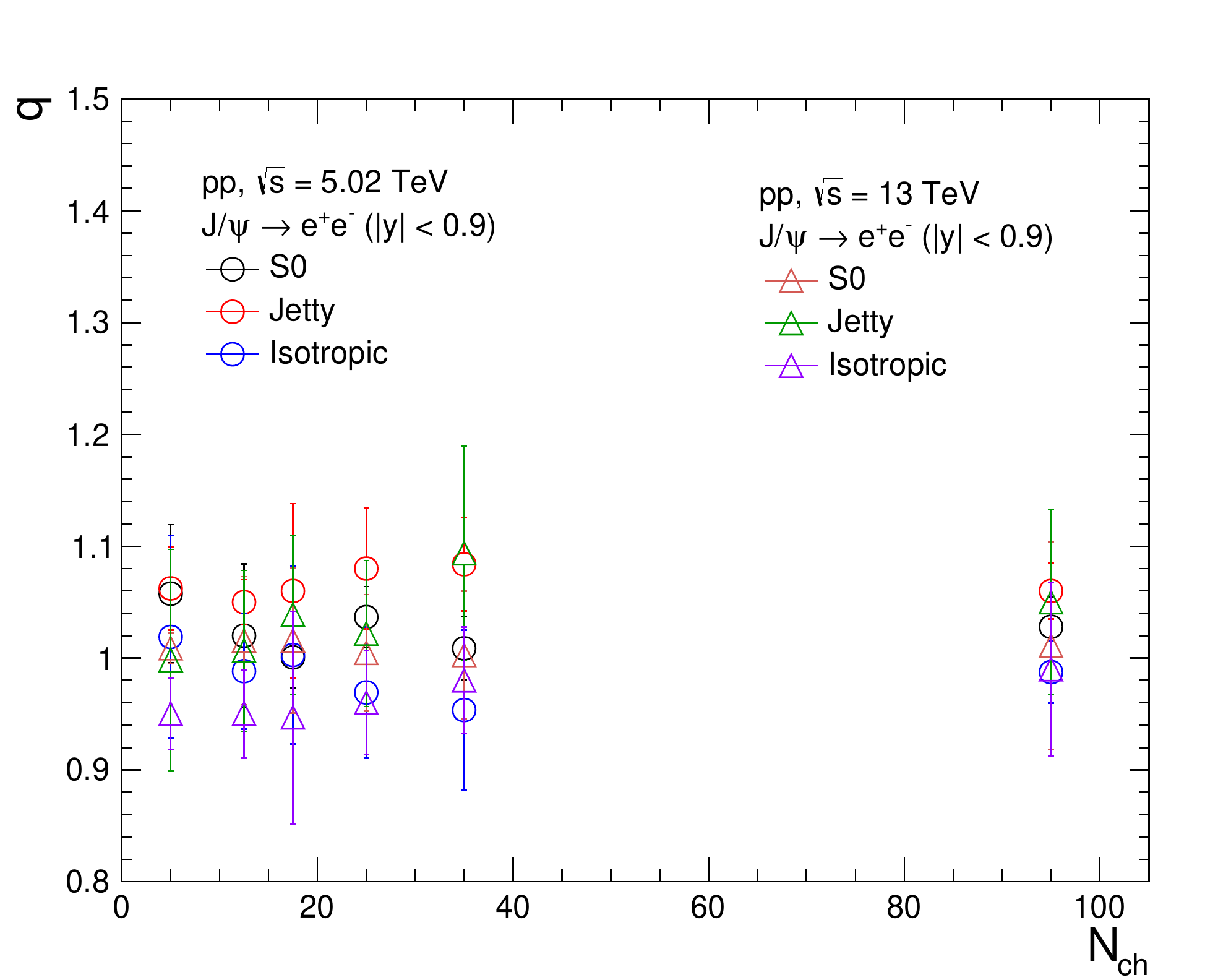}
\includegraphics[width=18pc]{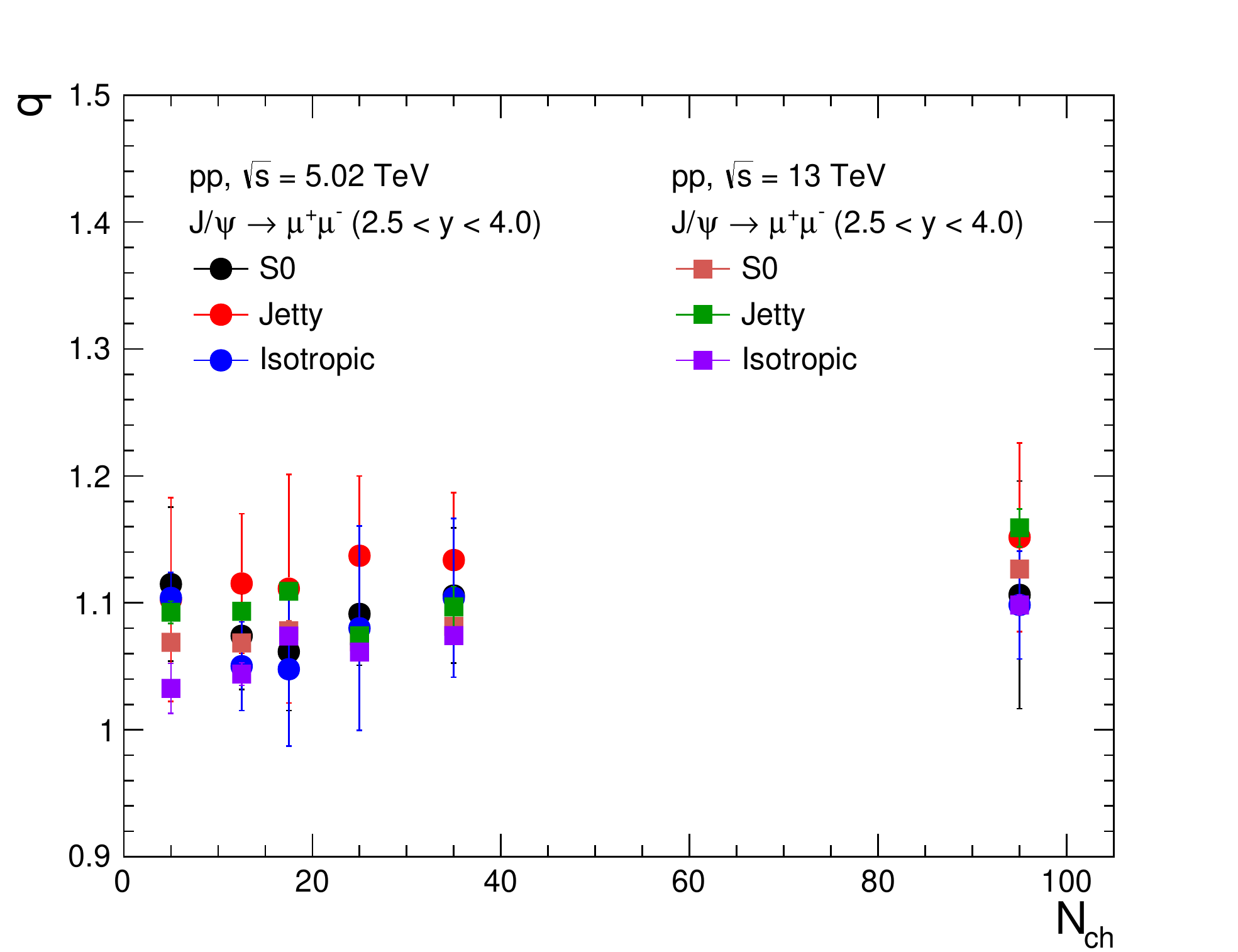}
\caption {(Color online) Multiplicity, rapidity and energy dependence of the non-extensive parameter ($q$) of the J/$\psi$ at mid and forward rapidities for pp collisions at $\sqrt{s}$ = 5.02 and13 TeV. The upper two panels represent the rapidity dependence and the lower two panels show energy dependence behavior.}
\label{fig:q}
\end{center}
\end{figure*}

The spectral description of Tsallis distribution function to J/$\psi$ production as a function of $p_T$ using PYTHIA8 are shown in Fig.\ref{fig7} for pp collisions at $\sqrt{s}$ = 5.02 TeV in the mid- and forward rapidities for jetty, isotropic and S$_{0}$-integrated events. The PYTHIA data to fit ratio is computed and is shown at the lower panel of Fig. \ref{fig7}, which shows all the points fall around unity except for the high $p_T$ bins where statistical fluctuations are larger. 
The analysis using Tsallis non-extensive statistics is repeated for different multiplicity classes and event shapes for pp collisions at $\sqrt{s}$ = 5.02 and 13 TeV. The ``q" and ``T" parameters have been studied as a function of multiplicity in three different event shape classes. Figure.~\ref{fig:Temp} and Fig.~\ref{fig:q} show the variation of Tallis-temperature and Tsallis non-extensive q-parameter as a function of multiplicity, rapidity and center-of-mass energy for three different event classes. 

A quantitative discussion of $\rm{J}/\psi$ production mechanism in non-extensive statistics is beyond the scope of this paper. We shall only discuss about the difference in the thermodynamics of $\rm{J}/\psi$ production at mid- and forward rapidities in jetty and isotropic events via their transverse momenta spectra. Important observations from the current study are described below:
 
 \begin{itemize}
 \item Tsallis-temperature for jetty and isotropic events shows an increasing trend with multiplicity irrespective of the rapidity under investigation
 \item Tsallis-temperature for J/$\psi$ is higher for higher center-of-energy ($T_{13} > T_{5.02}$)
 \item At the mid-rapidity, Tsallis-temperature of jetty events is higher than isotropic events whereas the behavior gets reversed at forward rapidity i.e $T_{isotropic} > T_{jetty}$
 \item Tsallis non-extensive q-parameter is consistent around 1.0$-$1.2, irrespective of center-of-mass energy, rapidity and multiplicity.
 \end{itemize}

As discussed in the introduction, the transverse momenta spectra of thermalized particles can be described by an exponentially decreasing behavior. But, at the higher momenta (at higher energies) the experimental data deviate from the usual Boltzmann function due to the dynamical effect of the system. The slope parameter represents the particle energy, which has both thermal (random) and collective contributions. The thermal motion gives the freeze out temperature ($T_{f}$), the temperature at which particles cease to interact with each other. In the presence of dynamical effect of a collective transverse flow, the increase of the slope parameter (T) at large $m_{T}$ ( $m_{T} = \sqrt{m^{2} + p_{T}^{2}}$ ) can be seen. Therefore, in the presence of collective transverse flow, $T = T_{f} +  m \langle v_{T} \rangle^{2}$,
where $\langle v_{T} \rangle$ is the average collective flow velocity and ``m" is the mass of the detected particle~\cite{Alberico:1999nh}.  The deviation from the Boltzmann slope at high-$p_{T}$ can be explained by the presence of non-extensive statistical effects \cite{Bhattacharyya:2015hya} , which incorporate the effect of a collective flow. The ``q" value extracted from the $p_{T}$ spectra of J/$\psi$ is not unity (shown in Fig.~\ref{fig:q} and Table~\ref{tab:q}), which reveals that the system contains dynamical effects and hence the presence of collective behaviour \cite{Bhattacharyya:2015hya,Wilk:2012zn}.  At the mid-rapidity, we can see from the Fig.~\ref{fig:Temp} and Table~\ref{tab:Temp} that the ``T" parameter  increases with an increase in multiplicity for both jetty and isotropic events. Further, the value of ``T" parameter for jetty events is higher than that of isotropic events. These observations support the statements that the collective-like effects increase from low to high-multiplicity irrespective of the effect of the event type in J/$\psi$ production. The effect is more prominent for jetty events compared to isotropic events. The observation goes in line with our prediction from ``crossing point" study that the collective effect may dominate towards higher multiplicities. When we look at the multiplicity dependent trend of ``T" parameter at forward rapidity (Fig.~\ref{fig:Temp}), the values are almost the same for different event multiplicities irrespective of the event shapes. This indicates that if collectivity is present in the system, it has almost equal effect irrespective of the multiplicity under investigation. The higher value of ``T" parameter in isotropic events with respect to the jetty events at forward rapidity might be due to the dominance of event type.  

\section{Summary}
  \label{sum}
 From the study of event shape using simulated events, multiplicity and rapidity dependence of J/$\psi$ production, we have drawn the following important observations:
  \newline
  \newline
 \textbf{Mid-rapidity:}
  \begin{itemize}
  \item  The jet contribution to the $\rm{J}/\psi$ production is larger at mid-rapidity compared to forward rapidity, and is independent of $\sqrt{s}$.
  
  \item The system formed in pp collisions contain dynamical effects, which leads to collective-like behaviour. The collectivity increases from low to high-multiplicity at mid-rapidity, irrespective of the  dominance of the event type in J/$\psi$ production, and is more prominent for jetty events compared to isotropic events.
   \end{itemize}
  
 \textbf{Forward rapidity:}
   \begin{itemize}
  \item The dominance of isotropic events is found throughout all the multiplicity bins at forward rapidities with a very little contribution from jetty events, and is independent of $\sqrt{s}$.
  
  \item From the study of Tsallis ``T" parameter at forward rapidity, we found the values are almost consistent with multiplicity for all the event types. Therefore, if collectivity is present in the system it has almost equal effect irrespective of the multiplicity under investigation.
   \end{itemize}
   
   The observation of completely different production dynamics of J/$\psi$ with multiplicity at the mid-rapidity and forward rapidity but independent of collision energy, supports the ALICE experimental results.

\section{Acknowledgement} 
DT acknowledges UGC, New Delhi, Government of India for financial supports. SD and RNS acknowledge the financial supports  from  ALICE  Project  No. SR/MF/PS-01/2014-IITI(G) and AK acknowledges the financial supports from  ALICE Upgrade project No. SR/MF/PS-01/2014-AMU of  Department  of  Science $\&$ Technology,  Government of India.  RNS acknowledges the financial supports from DAE-BRNS Project No. 58/14/29/2019-BRNS. This research used resources of the Grid computing facilities at the  Department of Physics, AMU, Aligarh, India.

  \begin {table*} [ht]
\scriptsize
\centering
\begin{tabular}{ |p{2.5cm}|p{2.2cm}|p{2.5cm}|p{2.2cm}|p{2.2cm}|p{2.2cm}|p{2.2cm}|p{2.5cm}|p{2.5cm}|p{2.5cm}|p{2.5cm}|p{2.5cm}|p{2.5cm}|  }
\hline
\multicolumn{7}{|c|}{Temperature (GeV) obtained in each multiplicity bin and event shape for $\sqrt{s}$ = 5.02 TeV} \\\hline
\multicolumn{1}{|c|}{}& \multicolumn{3}{|c|}{$|y|<$ 0.9}&\multicolumn{3}{|c|}{2.5$<y<$4.} \\\hline 
\bf{Mult-Bin}  &\bf{$S_{0}$} & \bf{Jetty} &\bf{Isotropy}&\bf{$S_{0}$} & \bf{Jetty} &\bf{Isotropy} \\\hline

 0 - 10   &0.597 $\pm$ 0.094&0.725$\pm$0.059&0.461$\pm$ 0.065&0.544 $\pm$ 0.059 & 0.425$\pm$0.031 &0.693 $\pm$ 0.049 \\\hline
  10 - 15   &0.769 $\pm$0.047&0.961$\pm$0.022&0.605$\pm$ 0.076&0.876 $\pm$ 0.043 & 0.488$\pm$ 0.033 &0.934 $\pm$ 0.064 \\\hline
   15 - 20   &0.859 $\pm$0.079&1.088$\pm$0.033&0.659$\pm$ 0.066&0.917 $\pm$ 0.047 &0.556$\pm$ 0.069&1.006 $\pm$ 0.080\\\hline
    20 - 30   &0.879 $\pm $0.074  &1.093$\pm$0.051&0.747$\pm$ 0.063&0.839$\pm$ 0.060&0.415$\pm$0.036&0.910 $\pm$ 0.076\\\hline
    30 - 40   &1.020 $\pm$0.092  &1.176$\pm$0.053&0.873$\pm$ 0.075& 0.836$\pm$ 0.073 &0.547$\pm$ 0.076&0.897 $\pm$ 0.080\\\hline
    40 - 150   &1.056 $\pm$0.073  &1.635$\pm$0.063&1.027$\pm$ 0.084& 0.831$\pm$ 0.075 &0.406$\pm$ 0.062&0.896 $\pm$ 0.081 \\\hline
  Integrated &0.799$\pm$ 0.021&0.878$\pm$ 0.067&0.715$\pm$ 0.006&0.807$\pm$ 0.021&0.797$\pm$ 0.064&0.762$\pm$ 0.043\\\hline
 \hline
   \multicolumn{7}{|c|}{Temperature (GeV) obtained in each multiplicity bin and event shape for $\sqrt{s}$ = 13 TeV} \\\hline  
    0-10           & 0.874 $\pm$0.054  &1.081 $\pm$ 0.088&0.761 $\pm$ 0.075& 1.097 $\pm$0.052& 0.876$\pm$0.090&1.433$\pm$0.197  \\\hline
    10 - 15        & 0.910 $\pm$ 0.049 &1.298 $\pm$ 0.078&0.853 $\pm$ 0.054&1.215$\pm$ 0.043 &1.015$\pm$0.010&1.395$\pm$ 0.075 \\\hline
    15 - 20       &0.971 $\pm$ 0.078 & 1.393 $\pm$ 0.093&0.949 $\pm$ 0.050&1.237 $\pm$  0.033  &1.021 $\pm$0.020&1.246$\pm$   0.012\\\hline
     20- 30       &1.085 $\pm$ 0.067 &1.695 $\pm$ 0.075&1.053 $\pm$ 0.065&1.430$\pm$ 0.035&1.376$\pm$    0.061  &1.488$\pm$ 0.067 \\\hline
     30 - 40      &1.173 $\pm$0.059  &1.700 $\pm$ 0.061&1.120 $\pm$ 0.073&1.391$\pm$  0.046 &1.196$\pm$  0.096  &1.462$\pm$ 0.073 \\\hline
     40 - 150   &1.200 $\pm$ 0.054 &1.771 $\pm$ 0.038 &1.200 $\pm$ 0.022&1.017$\pm$ 0.055 &0.850$\pm$ 0.072  &1.266$\pm$ 0.085 \\\hline
     Integrated &1.024 $\pm$ 0.095&1.072 $\pm$ 0.084&1.094 $\pm$ 0.097&1.181 $\pm$ 0.019&1.192$\pm$ 0.053&1.270$\pm$0.035 \\\hline 
   \end{tabular}
\caption{The extracted temperature parameters (T) from Tsallis distribution fitting ( Eq.~\ref{eq3} ) to the $p_{\rm{T}}$ spectra of J/$\psi$ along with statistical uncertainties in different multiplicity bins for pp collisions at $\sqrt{s}$ = 5.02 and 13 TeV.}
 \label{tab:Temp}
 \end {table*}

\begin {table*} [htb]
\scriptsize
\centering
 \begin{tabular}{ |p{2.5cm}|p{2.2cm}|p{2.5cm}|p{2.2cm}|p{2.2cm}|p{2.2cm}|p{2.2cm}|p{2.5cm}|p{2.5cm}|p{2.5cm}|p{2.5cm}|p{2.5cm}|p{2.5cm}|  }
 \hline
\multicolumn{7}{|c|}{Tsallis-q parameter obtained in each multiplicity bin and event shape for $\sqrt{s}$ = 5.02 TeV} \\\hline
 \multicolumn{1}{|c|}{}& \multicolumn{3}{|c|}{$|y|<$ 0.9}&\multicolumn{3}{|c|}{2.5$<y<$4.} \\\hline 
  \bf{Mult-Bin}  &\bf{$S_{0}$} & \bf{Jetty} &\bf{Isotropy}&\bf{$S_{0}$} & \bf{Jetty} &\bf{Isotropy} \\\hline
  
   0 - 10   &1.057$\pm$0.062&1.062$\pm$0.037&1.019$\pm$ 0.090 & 1.114 $\pm$0.061 & 1.102$\pm$ 0.080 &1.104$\pm$0.020\\\hline
  10 - 15   &1.020$\pm$0.064&1.050$\pm$0.020&0.988$\pm$ 0.051& 1.074 $\pm$0.042 & 1.115 $\pm$ 0.055 &1.050$\pm$0.035 \\\hline
   15 - 20   &1.000$\pm$0.028 &1.060$\pm$0.078&1.003$\pm$ 0.079&1.061 $\pm$ 0.046& 1.111 $\pm$ 0.090 &1.048$\pm$0.060 \\\hline
    20 - 30   &1.037$\pm$ 0.027&1.080$\pm$0.054&0.969$\pm$ 0.058 &1.091$\pm$ 0.040 &1.137 $\pm$ 0.063 &1.080$\pm$0.081 \\\hline
    30 - 40   &1.008$\pm$0.028 &1.084$\pm$0.042&0.953$\pm$ 0.071&1.106 $\pm$0.053&1.133 $\pm$ 0.053&1.104$\pm$ 0.062 \\\hline
    40 - 150   &1.028$\pm$0.027 &1.060$\pm$0.025&0.987$\pm$ 0.028& 1.106$\pm$0.090  &1.152 $\pm$ 0.074&1.098$\pm$0.042 \\\hline
    Integrated &1.033$\pm$ 0.004&1.073$\pm$ 0.014&1.020$\pm$ 0.001&1.089$\pm$ 0.002 &1.083$\pm$ 0.006 &1.097$\pm$ 0.004 \\\hline
   \hline
   \multicolumn{7}{|c|}{Tsallis-q parameter obtained in each multiplicity bin and event shape for $\sqrt{s}$ = 13 TeV} \\\hline
     0-10           & 1.010 $\pm$0.013  &0.998  $\pm$ 0.099&0.950 $\pm$ 0.032&1.069$\pm$ 0.005&1.092$\pm$ 0.008&1.033$\pm$ 0.019  \\\hline
    10 - 15        & 1.015 $\pm$0.057 &1.006  $\pm$ 0.072&0.950   $\pm$ 0.039&1.068$\pm$ 0.004&1.093$\pm$ 0.002&1.044$\pm$ 0.008 \\\hline
    15 - 20       &1.016  $\pm$ 0.065 & 1.040  $\pm$ 0.071&0.947  $\pm$ 0.095&1.078$\pm$ 0.007&1.109$\pm$ 0.002&1.074$\pm$ 0.002 \\\hline
     20- 30       &1.005  $\pm$ 0.052 &1.023  $\pm$ 0.065&0.960  $\pm$  0.046&1.069$\pm$ 0.003&1.074$\pm$ 0.002&1.061$\pm$ 0.006 \\\hline
     30 - 40      &1.003  $\pm$ 0.057  &1.094  $\pm$ 0.096&0.980 $\pm$  0.048&1.081$\pm$ 0.004&1.097$\pm$ 0.016&1.074$\pm$ 0.007 \\\hline
     40 - 150   &1.011  $\pm$ 0.093 &1.050  $\pm$ 0.083 &0.990  $\pm$   0.077&1.127$\pm$ 0.005&1.159$\pm$ 0.015&1.098$\pm$ 0.008 \\\hline
     Integrated &1.018$\pm$ 0.050&1.069$\pm$ 0.084&0.974$\pm$ 0.056&1.089 $\pm$0.002  &1.081$\pm$0.005&1.085$\pm$0.003\\\hline
   \end{tabular}
 \caption{The extracted non-extensive parameters (q) from Tsallis distribution fitting ( Eq.~\ref{eq3} ) to the $p_{\rm{T}}$ spectra of J/$\psi$ along with statistical uncertainties in different multiplicity bins for  pp collisions at $\sqrt{s}$ = 5.02 and 13 TeV. }
 \label{tab:q}
 \end {table*}

 \end{document}